\documentclass[preprint,aps,prd,groupedaddress,showpacs]{revtex4}
\usepackage{graphicx}
\usepackage{amsmath}
\usepackage{bm}
\usepackage{epsfig}

\usepackage{relsize}
\RequirePackage{xspace}

\begin{document}

\title{\mbox{}
Photoproduction of $J/\psi$ in association with a $c\bar{c}$ pair}

\author{Rong Li$~^{(a)}$ and Kuang-Ta Chao$~^{(a,b)}$}
\affiliation{ {\footnotesize(a)~ Department of Physics and State Key
Laboratory of Nuclear Physics and Technology, Peking University,
 Beijing 100871, China}\\
{\footnotesize (b)~Center for High Energy Physics, Peking
University, Beijing 100871, China}}

\begin{abstract}

Based on the color-singlet model, we investigate the photoproduction
of $J/\psi$ associated with a $c \bar{c}$ pair with all subprocesses
including the direct, single-resolved, and double-resolved channels.
The amplitude squared of these subprocesses are obtained
analytically. By choosing corresponding parameters, we give
theoretical predictions for the $J/\psi$ transverse momentum
distributions both at the LEPII and at the future photon colliders
for these subprocesses. The numerical results show that at the LEPII
these processes can not give enough contributions to account for the
experimental data, and it indicates that the color-octet mechanism
may still be needed. At the photon collider with the laser back
scattering photons, the resolved photon channe will dominate over
the direct one in small and moderate $p_t$ regions with large
$\sqrt{s}$. By measuring the $J/\psi$ production associated with a
$c\bar{c}$ pair, this process can be separated from the inclusive
$J/\psi$ production and may provide a new chance to test the
color-singlet contributions.

\end{abstract}

\maketitle

\section{Introduction}

Since the discovery of $J/\psi$, heavy quarkonium has provided an
ideal laboratory to investigate the fundamental  theory of strong
interactions, Quantum Chromodynamics (QCD). Conventionally, people
use the color-singlet model (CSM)~\cite{Einhorn:1975ua} to describe
the production and decay of heavy quarkonium. In order to overcome
the theoretical difficulties related to the infrared divergences in
the CSM~\cite{Barbieri:1980yp,Barbieri:1976fp} and  reconcile the
large discrepancy between the Tevatron data and the theoretical
prediction given by the CSM~\cite{Abe:1992ww}, an effective theory,
the non-relativistic quantum chromodynamics (NRQCD) factorization
formalism  was proposed~\cite{Bodwin:1994jh}.   In NRQCD the
production and decay rates of heavy quarkonium are factorized into
the short distance parts and the long distance parts, and because
the contributions of high Fock states are taken into account, the
intermediate $Q\bar{Q}$ pair that is produced in the short distance
part can be in various states with different angular momenta and
different colors. By introducing the color-octet mechanism (COM) in
NRQCD, one may resolve the problem of infrared divergences in the
CSM~\cite{Bodwin:1992ye} and may hope to give a proper
interpretation for the transverse momentum $p_t$ distribution of
$J/\psi$ production at the Tevatron~\cite{Braaten:1994vv}. More
detail descriptions on many aspects of heavy quarkonium physics can
be found in Ref.\cite{Brambilla:2004wf}.

The photoproduction of $J/\psi$ has been investigated by many
authors~\cite{Ma:1997bi,Klasen:2001mi,Klasen:2008mh}. In 2001 the
DELPHI Collaboration gave the measurement on inclusive
photoproduction of $J/\psi$~\cite{DELPHI}. Theoretical analysis
indicates that the $p_t$ distribution predicted in the CSM is an
order of magnitude smaller than the experimental result and the
NRQCD prediction can give a good account for it by the
COM~\cite{Klasen:2001cu}. It has been reviewed as a strong support
to the COM in NRQCD. The color evaporation model and the $k_t$
factorization formulism were also used to investigate this
process~\cite{Lipatov:2003es}. Furthermore, the
next-to-leading-order (NLO) QCD corrections to the processes $\gamma
+ \gamma \rightarrow c\bar{c}[{}^{3}S_1^{(8)}]+ g$ and $\gamma +
\gamma \rightarrow c\bar{c}[{}^{3}S_1^{(1)}]+ \gamma$ are
accomplished in~\cite{Klasen:2004tz} and the authors also give
theoretical predictions for these processes at the TESLA.

Recently, a number of studies show the importance of the heavy quark
pair associated $J/\psi$ production in the CSM. The contributions
from $J/\psi+c+\bar{c}$ final states in $J/\psi$ inclusive
production have been discussed by many authors at $B$
factories~\cite{Cho:1996cg,Zhang:2006ay} and LEP
~\cite{Qiao:2003ba}, and at the Tevatron and LHC
~\cite{Klasen:2008mh,Artoisenet:2007xi}, and even been studied in
the $k_t$ factorization formalism~\cite{Baranov:2006dh}. Although it
is a NLO process, the $p_t$ distribution can be changed and the
differential cross section can be enhanced at large $p_t$ due to the
different kinematics of the Feynman diagrams. At $B$ factories, the
$e^++e^- \to J/\psi+c+\bar{c}$ process gives more than half
contribution to the total cross section of the $J/\psi$ inclusive
production~\cite{Abe:2002rb}. In Ref.~\cite{Qiao:2003ba}, the
authors study the process $\gamma+\gamma \to J/\psi+c+\bar{c}$ and
find that the NLO process gives more contribution compared with that
of the leading-order (LO) process $\gamma+\gamma \to J/\psi+\gamma$
at the LEP. In the large $p_t$ region, the contribution from
$\gamma+\gamma \to J/\psi+c+\bar{c}$ is bigger than that of the
fragmentation process $\gamma+\gamma \to c+\bar{c}~frag. \to
J/\psi+c+\bar{c}$. In Ref.~\cite{Baranov:2006dh}, the process
$\gamma+g \to J/\psi+c+\bar{c}$ was studied in the $k_t$
factorization formalism.

In this paper, we will investigate all the subprocesses of the
photoproduction of $J/\psi$ associated with $c\bar{c}$ in the CSM.
Firstly, the full results including contributions from all the
single and double resolved photon processes of the $J/\psi$
production associated with heavy quark-antiquark pair at the LEP
will be presented for the first time. Secondly, these processes will
be extended to the photon colliders and the results with different
photon production mechanisms will be given.

The paper is organized as follows. In section II, we give
definitions of some relevant quantities and derive the analytical
formulas of the differential cross sections for all the
subprocesses. Numerical results are given in section III. Finally, a
summary is given in section IV.

\section{Formulation and calculation}
There are three classes of subprocesses for $\gamma + \gamma
\rightarrow J/\psi + c + \bar{c}+X$: As shown in Eq.~(\ref{direct}),
the direct process, where the two photons directly couple to the
final heavy quarks; In Eq.~(\ref{one-resolved}), the single-resolved
process, where one photon fluctuates to a parton (here, the gluon)
and collide with the other photon to produce the final states; In
Eq.~(\ref{double-resolved}), the double-resolved processes, where
both the two photons fluctuate to partons to produce the final
states. So in order to investigate the process thoroughly, the
following four subprocesses must be calculated:
\begin{eqnarray}
\label{direct}  \gamma + \gamma \rightarrow J/\psi + c + \bar{c}
\end{eqnarray}
\begin{eqnarray}
\label{one-resolved}  \gamma + g \rightarrow J/\psi + c + \bar{c}
\end{eqnarray}
\begin{eqnarray}
\label{double-resolved} \nonumber g + g \rightarrow J/\psi + c + \bar{c} \\
  q + \bar{q} \rightarrow J/\psi + c +
\bar{c}
\end{eqnarray}
\begin{figure}[!htbp]
 \begin{center}
  \includegraphics[width=0.90\textwidth]{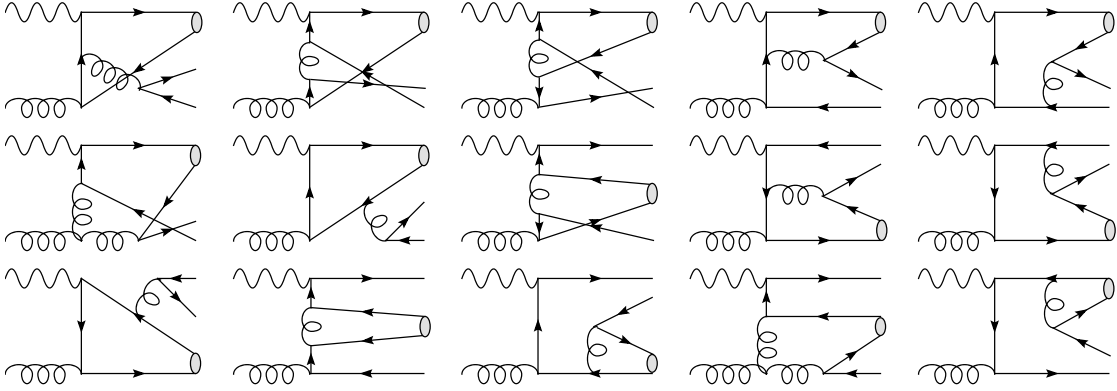}
    \caption{Typical Feynman diagrams for subprocesses
     $\gamma+g \to J/\psi+ c+ \bar{c}$. The others can be obtained by reversing the fermion lines.}
   \label{fig:gagdiagram}
 \end{center}
\end{figure}

\begin{figure}[!htbp]
 \begin{center}
  \includegraphics[width=0.90\textwidth]{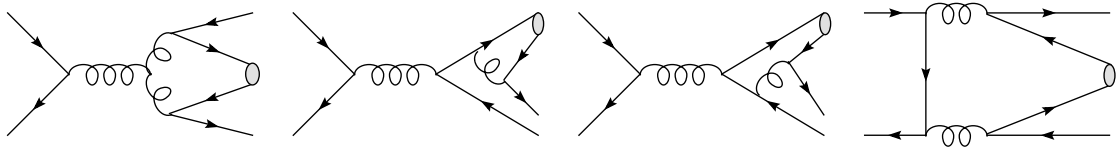}
    \caption{Typical Feynman diagrams for subprocesses
     $q+\bar{q}\to J/\psi+c+\bar{c}$. The others can be obtained by reversing the fermion lines.}
   \label{fig:qqdiagram}
 \end{center}
\end{figure}
The four subprocesses involve 20, 30, 42, 7 Feynman diagrams,
respectively. Fig.\ref{fig:gagdiagram} and \ref{fig:qqdiagram} just
show the Feynman diagrams of the processes $\gamma + g \rightarrow
J/\psi + c + \bar{c}$ and $q +\bar{q}  \rightarrow J/\psi + c +
\bar{c}$. The Feynman diagrams of the other two subprocesses are as
same as those given in the Ref.~\cite{Qiao:2003ba,Artoisenet:2007xi}. Following the
color-singlet factorization formalism and the standard covariant
projection method~\cite{Baier:1983va}, the scattering amplitudes of these
subprocesses can be expressed as
\begin{eqnarray}
\label{amp2}   &&\hspace{-1cm}{\cal M}(a(k_1)+b(k_2)\rightarrow
c\bar{c}(^{2S+1}L_{J}^{(1)})(P)+c(p_1)+\bar{c}(p_2))=\sqrt{C_{L}}
\sum\limits_{L_{z} S_{z} }\sum\limits_{s_1s_2 }\sum\limits_{jk}
\nonumber
\\ & \times&\langle s_1;s_2\mid
S S_{z}\rangle \langle L L_{z};S S_{z}\mid J J_{z}\rangle\langle
3j;\bar{3}k\mid 1\rangle\nonumber\\
&\times&
\begin{array}{ll}
{\cal M}(a(k_1)+b(k_2)\rightarrow
 c_j(\frac{P}{2};s_1)+\bar{c}_k(\frac{P}{2};s_2)+c(p_1)+\bar{c}(p_2)),
\end{array}
\end{eqnarray}
where $\langle 3j;\bar{3}k\mid 1\rangle$, $\langle s_1;s_2\mid S
S_{z}\rangle$ and $\langle L L_{z};S S_{z}\mid J J_{z}\rangle$ are
the color-SU(3), spin-SU(2), and orbital angular momentum
Clebsch-Gordan coefficients respectively for $c\bar{c}$ pairs
projecting out appropriate quantum numbers of the bound states. The
$C_{L}$ is the probability that describes a heavy quark-antiquark
pair having the appropriate quantum numbers to evolved into a
corresponding meson and, for the $J/\psi$ here, can be related to
the wave function at the origin $R(0)$ or the color-singlet long
distance matrix element $\langle0|O^{J/\psi}_1|0\rangle$ as
following
\begin{equation}
    \begin{array}{rcl}
C_L&=& \frac{1}{4\pi}|R(0)|^2\nonumber\\
&=& \frac{1}{2N_c(2J+1)}\langle0|O^{J/\psi}_1|0\rangle.
    \end{array}
\end{equation}

As for $J/\psi$ production, the spin-triplet projection operator
should be used, which is defined as
\begin{equation}
    \begin{array}{rcl}
P_{1S_Z}(P,0)&=& \sum\limits_{\frac{1}{2} \frac{1}{2}}\langle
\frac{1}{2};\frac{1}{2}\mid 1 S_{z}\rangle
v(\frac{P}{2};\frac{1}{2})\bar{u}(\frac{P}{2};\frac{1}{2})\nonumber\\
&=& \frac{1}{2\sqrt{2}}\not{\epsilon}(S_z)(\not{P}+2m_{c}).
    \end{array}
\end{equation}
At the same time, the color projection operator for the
color-singlet state is given by

\begin{equation}
\langle 3j;\bar{3}k\mid 1\rangle=\delta_{ij}/\sqrt{N_c}.
\end{equation}

We use the FeynArts~\cite{Kublbeck:1990xc} to generate the Feynman
diagrams and amplitudes in the Feynman gauge, then insert the
projection operators and use the FeynCalc~\cite{Mertig:1990an} to
evaluate the square of the amplitudes. In calculating the
subprocesses $g+g \to J/\psi +c+\bar{c}$,  $-g^{\mu\nu}$ is used for
the polarization summation of the initial gluons and therefore the
corresponding contribution of the ghost diagrams must be subtracted.
The analytical results for every sbuprocesses are too tedious to be
shown in this paper. In order to check the gauge invariance, the
polarization vector of one initial gluon (photon) is replaced by the
corresponding momentum in the direct and single-resolved processes
and the zero results are obtained at the level of squared matrix
element analytically. To check the gauge invariance of the
subprocess $g+g \to J/\psi +c+\bar{c}$, we replace the polarization
vector of one of the initial gluons by its momentum and use the
physical polarization tensor $P_{\mu\nu}$ for the polarization
summation of the other gluon. Then the square of the amplitude
vanishes. Otherwise, the ghost diagrams must be taken into
consideration for checking the gauge invariance. Here the physical
polarization tensor $P_{\mu\nu}$ is explicitly expressed as
\begin{equation}
\label{sumgluon}
P_{\mu\nu}=-g_{\mu\nu}+\frac{k_\mu\eta_\nu+k_\nu\eta_\mu}{k\cdot\eta},
\end{equation}
where k is the momentum of the gluon, $\eta$ is an arbitrary
light-like four vector with $k \cdot \eta \ne 0$. In the
calculation, $\eta$ is set as the momentum of the other initial
gluon conveniently.

The differential cross section can be obtained by convoluting the
parton level differential cross section with the photon density
functions and the parton distribution functions of the photon. It is
expressed as
\begin{eqnarray}
\label{xs} d\sigma(e^++e^- \rightarrow e^++e^-+
J/\psi+c+\bar{c})=\int dx_1 dx_2 f_{\gamma}(x_1)
f_{\gamma}(x_2)\nonumber
\\  \times\sum \limits_{i,j} \int dx_i dx_j f_{i/\gamma}(x_i)
f_{j/\gamma}(x_j) d\hat{\sigma}(i + j \rightarrow J/\psi+c+\bar{c})
,
\end{eqnarray}
where $f_{\gamma}(x)$ is the photon density function and
$f_{i/\gamma}(x)$ is the parton distribution function of the photon.
Here the labels i and j denote the parton contents of the photon,
such as gluon and the light quarks. In the direct photon process,
the distribution function $f_{\gamma/\gamma}(x)=\delta(1-x)$.

In the photon-photon collisions, the initial photons can be
generated by the bremsstrahlung or by the laser back scattering
(LBS) from the $e^+e^-$ collision. The spectrum of the
bremsstrahlung photon can be described by the Weizsacker-Williams
approximation (WWA) as following~\cite{Williams:1934ad}
\begin{eqnarray}
\label{WWA} f_{\gamma}(x)=\frac{\alpha}{2 \pi}\left(2
m_e^2(\frac{1}{Q_{max}^2}-
\frac{1}{Q_{min}^2})x+\frac{(1+(1-x)^2)}{x}\log(\frac{Q_{max}^2}{Q_{min}^2})\right),
\end{eqnarray}
where $x=E_{\gamma}/E_e$, $\alpha$ is the the fine structure
constant and $m_e$ is the electron mass. The definition of
$Q_{max}^2$ and $Q_{min}^2$ are given by
\begin{eqnarray}
Q_{min}^2&=&\frac{m_e^2 x^2}{1-x},\\
Q_{max}^2&=&(\frac{\sqrt{s} \theta}{2})^2 (1-x)+Q_{min}^2,
\end{eqnarray}
where $\theta$ is the angle between the momentum of the photon and
the direction of the electron beam. This angle is taken as 32mrad at
the LEPII. On the other hand, the laser back scattering can generate
more energetic and luminous photons. The spectrum of the LBS photon
is expressed as~\cite{Ginzburg:1981vm}
\begin{eqnarray}
\label{LBS} f_\gamma(x) &=&  \frac{1}{N}\left[1 - x + \frac{1}{1 -
x} - 4 r (1 - r)\right],
\end{eqnarray}
where $x=E_{\gamma}/E_e$, $r=\frac{x}{x_m(1-x)}$, and the constant N
is given by
\begin{eqnarray}
 N =  \left(1 - \frac{4}{x_m} -
\frac{8}{x_m^2}\right)\log(1 + x_m) + \frac{1}{2} + \frac{8}{x_m} -
\frac{1}{2 (1 + m_x)^2},
\end{eqnarray}
where $x_m=4E_bE_l\cos^2\frac{\theta}{2}$. Here $E_b$ is the energy
of electron beam, $E_l$ is the energy of the incident laser beam,
$\theta$ is the angle between the laser and the electron beam. The
energy of the LBS photon is restricted by the following equation
\begin{eqnarray}
0 \le x \le \frac{x_m}{1 + x_m}~,
\end{eqnarray}
Telnov~\cite{Telnov:1989sd} argued that the optimal value of $x_m$
is 4.83.

\begin{figure}[!htbp]
 \begin{center}
  \includegraphics[width=0.90\textwidth]{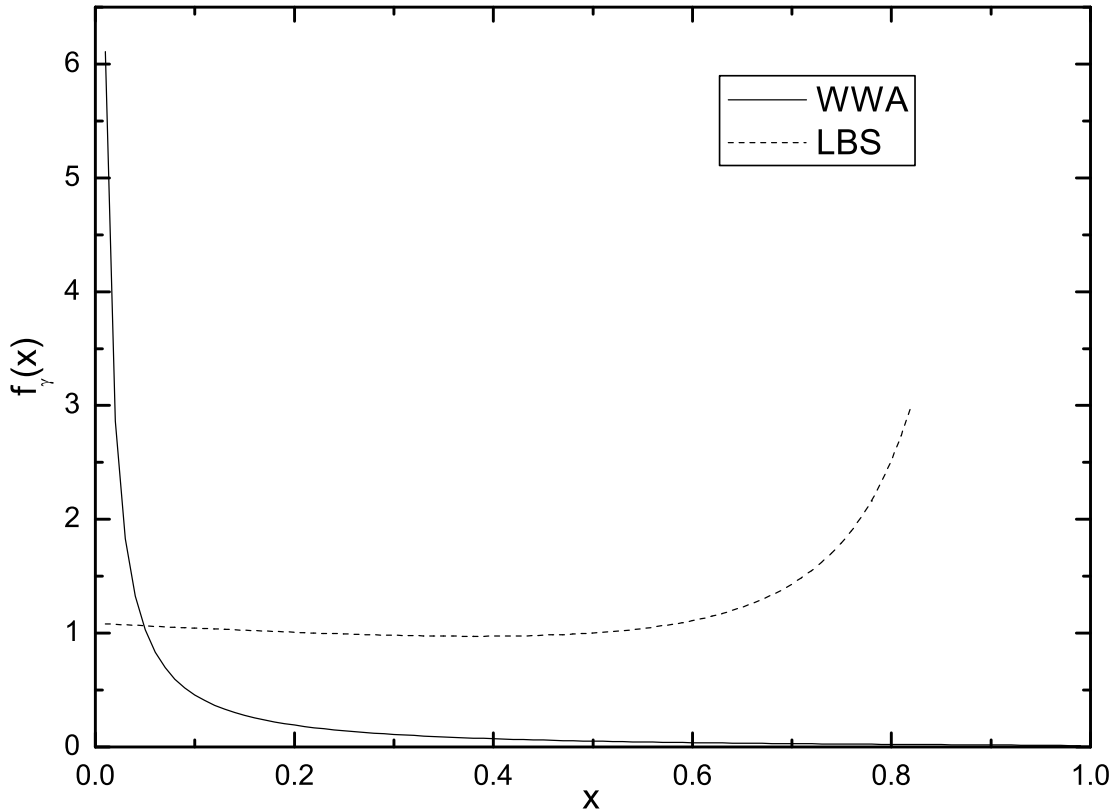}
    \caption{The photon spectra of the WWA and LBS at $\sqrt{s}=500$GeV.}
   \label{fig:gamma-pdf}
 \end{center}
\end{figure}

The spectra of the LBS and WWA photons are very different. While the
latter depends only on the center-of-mass energy, the former depends
on the parameter $x_m$ also. By comparing the spectra of the WWA
photon at $\sqrt{s}=500$ GeV to the one at $\sqrt{s}=1$ TeV, we
clearly see that there is no qualitative difference between them and
the numerical difference is less than $15\%$. Therefore, we just
show the comparison of the the spectra of WWA photon and that of the
LBS photon at $\sqrt{s}=500$ GeV in the Fig.~\ref{fig:gamma-pdf}.
And it can be seen that the distribution of the WWA photon is large
at the small x region and tends to infinite at the end-point
$x\approx0$. On the contrary, the distribution of the LBS photon is
moderate in the whole x region and get its maximum value at the
largest x point. These two distributions can result in significant
different results.

\section{Numerical results and discussions}

In calculating the numerical results, we choose the following
parameters: $M_c=1.5$ GeV, $\alpha=1/137$, $m_e=0.511$ MeV and the
color-singlet matrix element ${\langle 0 |} {\cal
O}^{J/\psi}({}^3S_1^{[1]}){| 0 \rangle}=1.4$
GeV$^3$~\cite{Qiao:2003ba}. The GRS99~\cite{Gluck:1999ub} parton
distribution function of photon is used and the running of
$\alpha_s$ is evaluated by the LO formula of
GRV98~\cite{Gluck:1998xa}. Both the renormalization and
factorization scales are fixed as $\sqrt{4M_c^2+p_t^2}$. The
numerical results are multiplied by a factor of 1.278 to include the
feeddown contribution from the $\psi'$~\cite{Qiao:2003ba}.

\begin{figure}[!htbp]
 \begin{center}
  \includegraphics[width=0.90\textwidth]{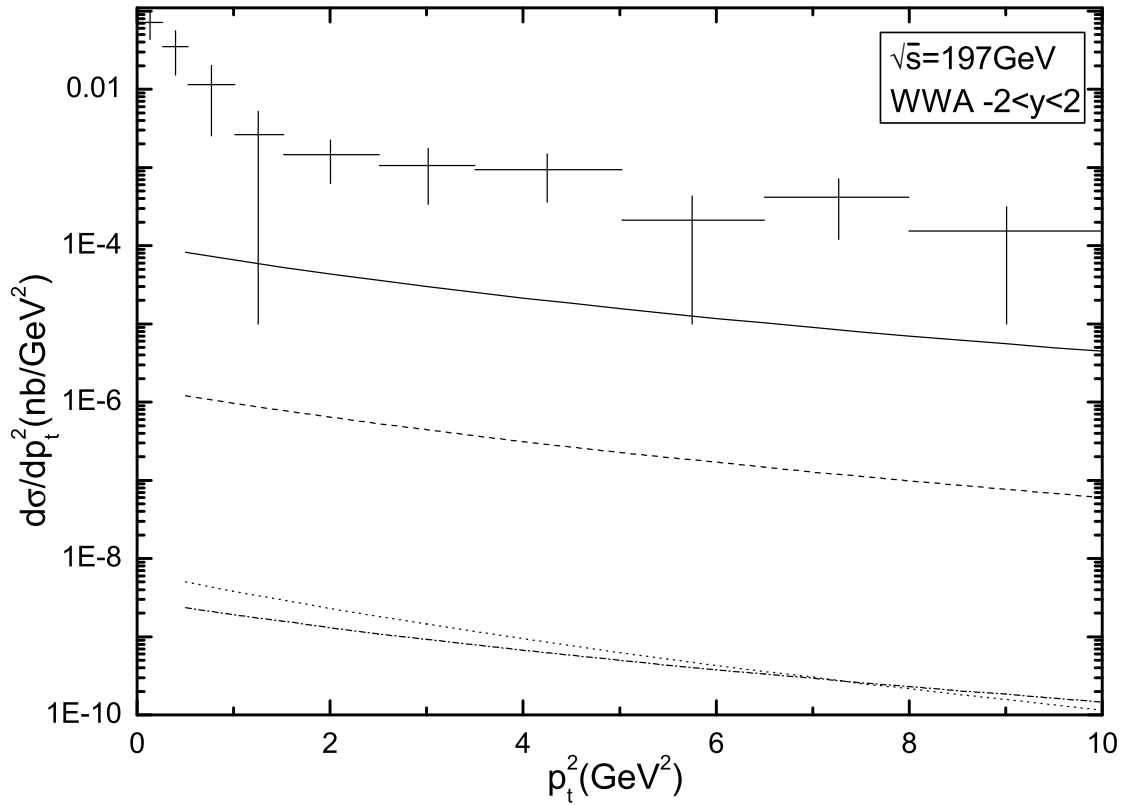}
  \caption{ $p_t$ distributions of differential cross sections of $J/\psi + c + \bar{c}$ production in various subprocesses at the LEPII.
  The solid, dashed, dotted and dash-dotted lines correspond
  to the subprocesses $\gamma+\gamma \to J/\psi
  +c+\bar{c}$, $\gamma+g \to J/\psi +c+\bar{c}$, $g+g \to J/\psi +c+\bar{c}$
  and $q+\bar{q} \to J/\psi +c+\bar{c}$, respectively. The experimental result of DELPHI is also presented\cite{DELPHI}.}
   \label{fig:LEPII}
 \end{center}
\end{figure}

The differential cross sections $d\sigma/dp_t^2$ for all the four
subprocesses at the LEPII are shown in Fig.\ref{fig:LEPII}. To
obtain theoretical predictions, the parameters which are related to
the LEPII experimental conditions are chosen as $\sqrt{s}=197$ GeV,
$\theta_{max}=32$ mrad and the rapidity cut $-2<y<2$. The constraint
of center-of-mass energy for the two photons is $W\leq 35$
GeV\cite{DELPHI}. From Fig.\ref{fig:LEPII}, one can see that the
direct photon subprocess is dominant at the LEPII with the WWA
photon. The contribution from the single-resolved subprocess is
smaller than that of the direct one by an order or more in magnitude
and the contributions from the double-resolved subprocesses are even
smaller than that of the direct one by almost four orders in
magnitude. In the $p_t$ region that we investigated, the
contribution from the double-resolved gluon subprocess and
quark-antiquark subprocess are of the same order in magnitude.

\begin{figure}
\begin{tabular}{cc}
\epsfig{file=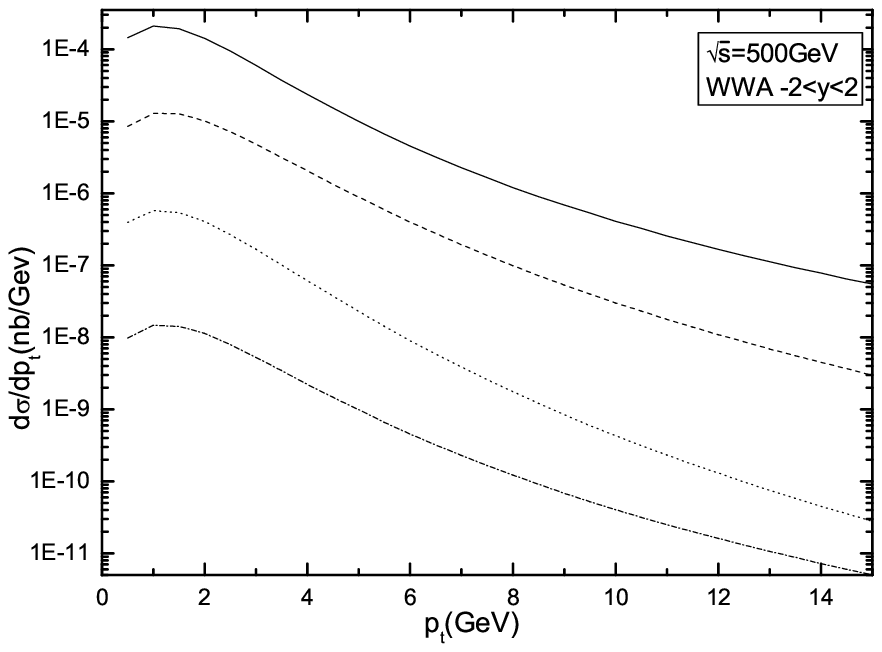,width=7cm}&
\epsfig{file=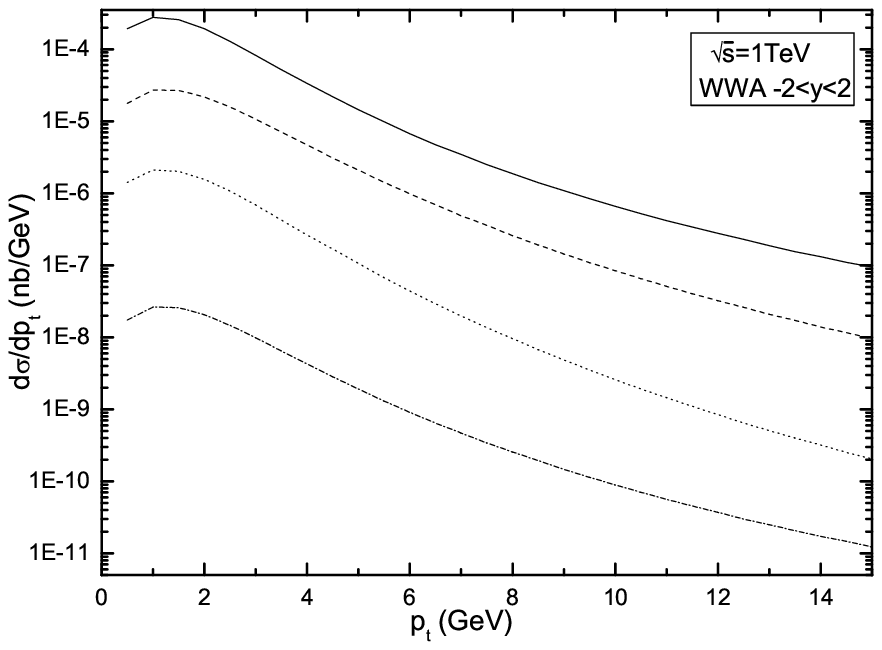,width=7cm}\\
\end{tabular}
\caption{$p_t$ distributions of differential cross sections of
$J/\psi + c + \bar{c}$ production at the photon collider with the
WWA photon spectrum at different $\sqrt{s}$. Here we use the same
notations as those in Fig.\ref{fig:LEPII}.} \label{fig:gamma-WWA}
\end{figure}

\begin{figure}
\begin{tabular}{cc}
\epsfig{file=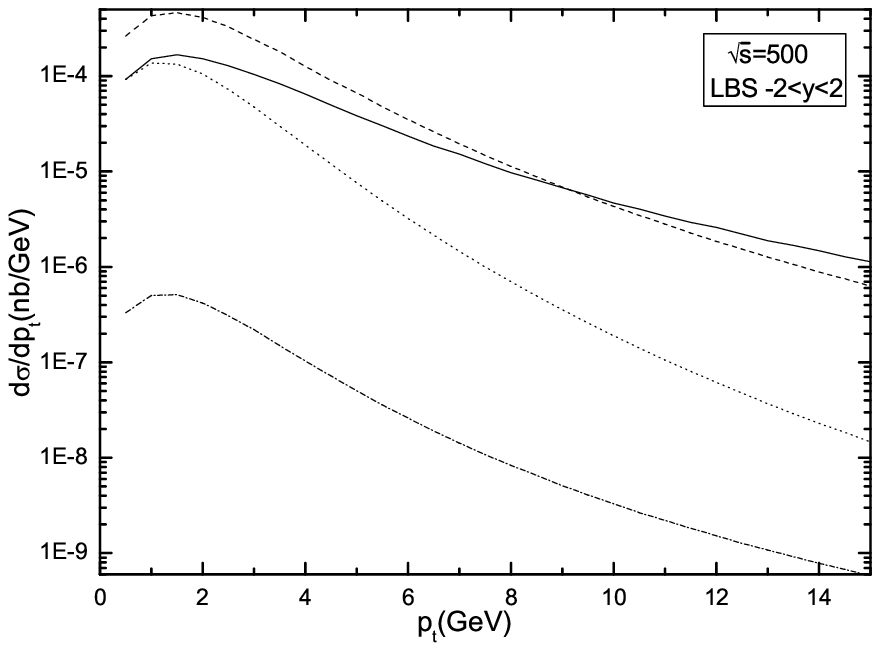,width=7cm}&
\epsfig{file=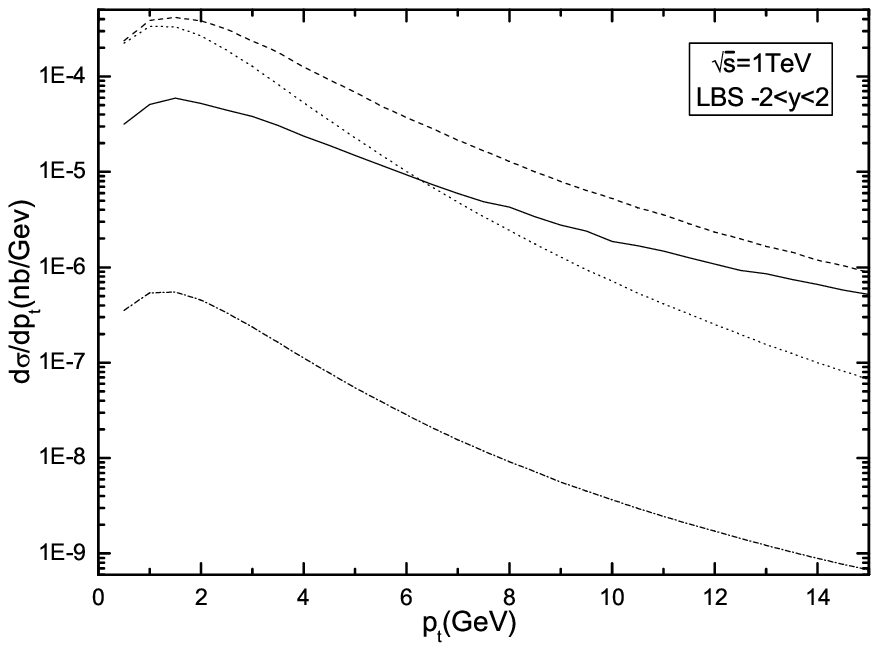,width=7cm}\\
\end{tabular}
\caption{$p_t$ distributions of differential cross sections of
$J/\psi + c + \bar{c}$ production at the photon collider with the
LBS photon spectrum at different $\sqrt{s}$. Here we use the same
notations as those in Fig.\ref{fig:LEPII}.} \label{fig:gamma-LBS}
\end{figure}

In the future, the $e^+e^-$ collider may run at $\sqrt{s}=500$GeV or
even at $\sqrt{s}=1$ TeV, and the LBS photon collision may be
realized. Therefore, we also investigate the four subprocesses at
$\sqrt{s}=500$ GeV and $\sqrt{s}=1$ TeV. For comparison, we give the
theoretical results with both the WWA photon and LBS photon at these
two center-of-mass energy. Here we set the $\theta_{max}$ of WW
approximation as 20mrad~\cite{Klasen:2001mi} and the $x_m$ of LBS
photon as 4.83, which determines the maximum photon energy fraction
as 0.83~\cite{Telnov:1989sd}. In contrast to the calculation for the
LEPII, here we do not use the constrain $W\leq 35$ GeV.

Fig.\ref{fig:gamma-WWA} and Fig.\ref{fig:gamma-LBS} give the $p_t$
distributions of the differential cross sections at different
center-of-mass energies with the LBS photon and WWA photon,
respectively, at photon collider. For the WWA photon case, the
direct photon production subprocess is always the dominant one. The
contribution from the single-resolved process is less than that from
the direct one, but larger than those from the double-resolved
processes. However, in the case of the LBS photon, with the increase
of the center-of-mass energy, the contributions from the
single-resolved and the double-resolved gluon subprocesses are
compatible with or even larger than that from the direct one. But
the contribution from the quark-antiquark subprocess is much smaller
than those from the other three subprocesses.

\begin{figure}[!htbp]
 \begin{center}
  \includegraphics[width=0.90\textwidth]{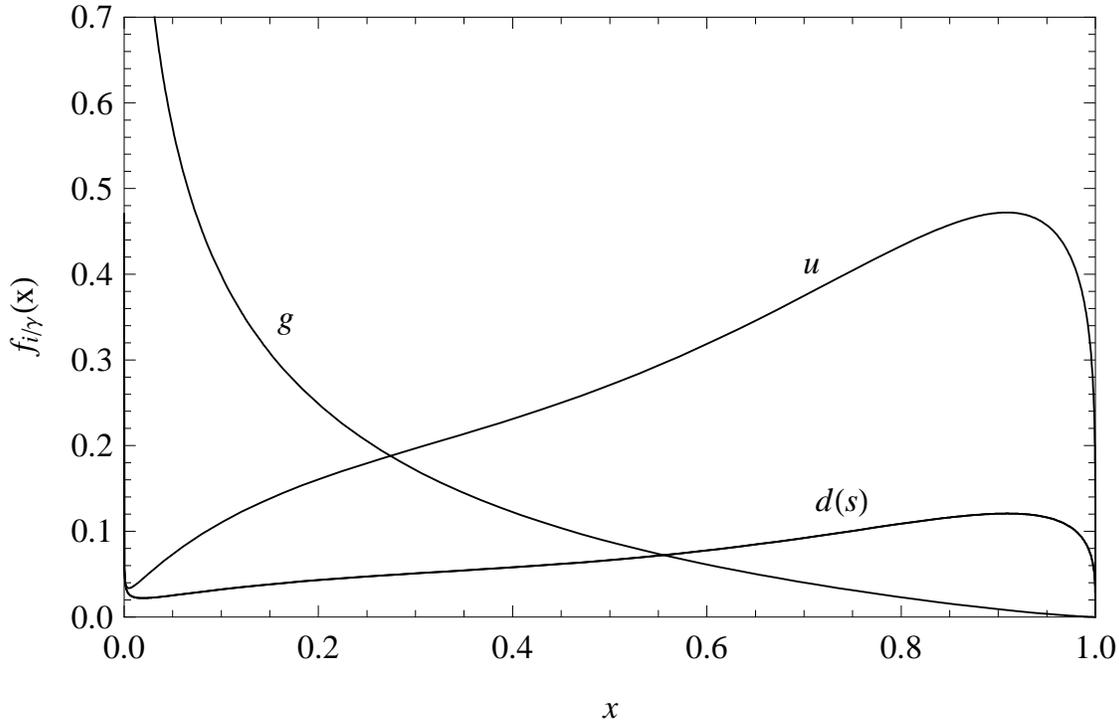}
    \caption{Parton distributions of the photon in the GRS99~\cite{Gluck:1999ub} parametrization
     at $Q^2=30$ GeV$^2$. Here x is the parton energy fraction.}
   \label{fig:GRS99}
 \end{center}
\end{figure}

Fig.\ref{fig:GRS99} shows the parton distributions of photon in the
GRS99 parametrization~\cite{Gluck:1999ub}. It can be seen that the
gluon content is dominant in small x region and even divergent when
x tends to zero. It is only in the large x region that the quark
contents can be dominant.

Let us first consider the subprocesses with the LBS photons as
initial states. When the $p_t$ of $J/\psi$ is lower or the
$\sqrt{s}$ becomes larger, the contributions from small x region
partons are dominant. Because the LBS photon spectrum function has
no singularity at small x region, and at the same time the gluon
distribution function of photon has a great enhancement at small x
region, the single-resolved and double-resolved gluon subprocess can
be dominant in the $J/\psi$ production with lower $p_t$ or larger
$\sqrt{s}$. However, as for the subprocesses with the WWA photons as
initial states, where both the photon spectrum function and the
gluon distribution function of the photon have great enhancements at
small x region, the single- and double-resolved subprocesses have no
predominance compared with the direct one in all the region of
$p_t$.

\begin{table}
\caption[]{Integrated cross sections of photoproduction of $J/\psi$
associated with a $c\bar{c}$ pair for different initial state
$e^+e^-$ energy $\sqrt{s}$ and subprocesses.  The $p_t$ cut of
$J/\psi$ is set as $p_t>$1 GeV. Other parameters and cut conditions
are chosen as the one being used to calculate the $p_t$
distributions in the text. (units: $\sqrt{s}$ in GeV and $\sigma$ in
nb.)} \label{tab:sigma}
\begin{center}
\renewcommand{\arraystretch}{1.5}
\[
\begin{array}{|c|c|c|c|c|}
\hline\hline
 \sqrt{s} & \sigma_{\gamma+\gamma} & \sigma_{\gamma+g} & \sigma_{g+g} & \sigma_{q+\bar{q}} \\ \hline
 197(WWA)& 2.06\times10^{-4} & 2.91\times10^{-6} & 8.68\times10^{-9} & 6.38\times10^{-9} \\ \hline
 500(WWA)& 3.51\times10^{-4} & 2.53\times10^{-5} & 9.74\times10^{-7} & 2.83\times10^{-8} \\ \hline
 500(LBS)& 5.33\times10^{-4} & 1.17\times10^{-3} & 2.57\times10^{-4} & 1.28\times10^{-5} \\ \hline
 1000(WWA)& 4.80\times10^{-4} & 5.54\times10^{-5} & 3.80\times10^{-6} & 5.22\times10^{-8} \\ \hline
 1000(LBS)& 1.97\times10^{-4} & 1.13\times10^{-3} & 6.66\times10^{-4} & 1.21\times10^{-6} \\ \hline \hline
\end{array}
\]
\renewcommand{\arraystretch}{1.0}
\end{center}
\end{table}

Table ~\ref{tab:sigma}. gives the integrated cross sections of every
subprocesses of the photoproduction of $J/\psi$ associated with
$c\bar{c}$ pair. From the numerical results, it can be seen that the
total contributions from the resolved (including single and double
resolved) subprocesses are smaller than that of direct one by about
an order in magnitude in the case of WWA photons. On the contrary,
in the case of LBS photons the total contributions from the resolved
subprocesses are larger than that from the direct one for by a
factor of 2.5 at $\sqrt{s}=500$ GeV and 9 at $\sqrt{s}=1000$ GeV. It
can also be inferred from the $p_t$ distribution presented in
Fig.~\ref{fig:LEPII},~\ref{fig:gamma-WWA} and \ref{fig:gamma-LBS}.
At the same time, all the integrated cross sections increase with
the increase of $\sqrt{s}$ for the processes initiated by the WWA
photons. And in the case of LBS photons, only the cross section of
the subprocess $g+g\to J/\psi+c\bar{c}$ increased with $\sqrt{s}$
enhanced from 500GeV to 1TeV.

For the direct photon subprocess, our numerical result is a little
different from the one in Ref.~\cite{Qiao:2003ba}. The numerical
results indicate that the contributions from the single-resolved and
double-resolved processes are much less than that from the direct
one at the LEPII. The authors of Ref.~\cite{Klasen:2004tz} have
given the results of the NLO QCD corrections for the subprocesses
$\gamma+\gamma \to J/\psi+\gamma$ and $\gamma+\gamma \to
c\bar{c}[^3S^1_8]+g$ at the TESLA. For the subprocess $\gamma+\gamma
\to J/\psi+\gamma$, the $K$ factor is smaller than one. And the QCD
correction for the color-octet subprocess $\gamma+\gamma \to
c\bar{c}[^3S^1_8]+g$ can enhanced the differential cross section
significantly in the large $p_t$ region. From the above NLO results
at the TESLA, one can expect that the NLO corrections to the
color-singlet subprocess $\gamma+\gamma \to J/\psi+\gamma$ could not
enhance the result largely at the LEPII also. So the contributions
from the color-octet mechanism can not be excluded in the inclusive
$J/\psi$ photoproduction at the LEPII. The full investigation on the
NLO QCD radiative corrections on the direct and resolved
subprocesses may help us to clarify the situation.

As for the photon collider with LBS initial photons, the
contributions from the single- and double-resolved photon
subprocesses become large significantly at lower and moderate $p_t$
region with large $\sqrt{s}$. This feature comes from the small x
behavior of the gluon content distribution function of the photon,
and can be checked in the future.

\section{Summary}
In this paper, we investigate the production of $J/\psi$ associated
with a $c\bar c$ pair in the CSM in photon-photon collisions,
including the direct, single-resolved and double-resolved
subprocesses. The formulas for the cross sections of the four
subprocesses are obtained in the collinear factorization formulism.
Moreover, the results of the single-resolved subprocess are given
for the first time. The numerical results show that the
contributions from color-octet processes can not be excluded at
present with the LEP experiment.

At the photon collider with the LBS initial photons, the
single-resolved and even the double-resolved processes will dominate
over the direct one in the small and moderate $p_t$ regions. By
measuring the final state $J/\psi$ and $c\bar{c}$ pair, the process
$\gamma+\gamma \to J/\psi +c+\bar{c}+X$ can be separated from the
inclusive $J/\psi$ production and could provide a channel to probe
the parton contents of the photon. Furthermore, to separate this
channel in experiment and compare the data with the theoretical
prediction in the CSM also gives a new chance to test the CSM
contributions.

\section*{Acknowledgments}
We thank Prof. Jian-Xiong Wang, Dr. Ce Meng and Dr. Yu-Jie Zhang for
helpful discussions. R.L. also thanks Dr. E. Reya and Dr. I.
Schienbein for providing the fortran code on the GRS parton
distribution of the photon. This work was supported by the National
Natural Science Foundation of China (No 10675003, No 10721063).

\newpage

\end{document}